\documentclass[pra,superscriptaddress,a4paper,twocolumn]{revtex4}
\usepackage{geometry}
\usepackage[english]{babel}
\usepackage[latin1]{inputenc}
\usepackage{hyperref}
\usepackage{amsmath, amssymb, amsfonts, mathrsfs}
\usepackage{amsthm}
\usepackage{graphicx}
\usepackage{dsfont}

\graphicspath{{images/}}
\usepackage{xcolor}
\usepackage{exscale}
\usepackage{graphicx}
\usepackage{amsmath}
\usepackage{latexsym}
\usepackage{amsfonts}
\usepackage{amssymb}
\usepackage{bbm}

\DeclareMathOperator{\Tr}{Tr}

\def\pmx{\begin{pmatrix}}
\def\emx{\end{pmatrix}}

\newtheorem{theorem}{Theorem}

\newtheorem{cor}{Corollary}

\newtheorem{definition}{Definition}
\newtheorem{theorem2}{Theorem}[section]
\newtheorem{lemma2}{Lemma}[section]

\newtheorem{cor2}{Corollary}[section]
\newtheorem{obs2}{Observation}[section]

\begin{document} 

\title{Characterizing multipartite entanglement without shared reference frames}
\author{C. Kl\"ockl}
\affiliation{Universitat Autonoma de Barcelona, 08193 Bellaterra, Barcelona, Spain}
\affiliation{Institute for Quantum Optics and Quantum Information (IQOQI), Austrian Academy of Sciences, Boltzmanngasse 3, 1090 Vienna, Austria}
\author{M. Huber}
\affiliation{Universitat Autonoma de Barcelona, 08193 Bellaterra, Barcelona, Spain}
\affiliation{ICFO-Institut de Ciencies Fotoniques, 08860 Castelldefels, Barcelona, Spain}

\begin{abstract}
Multipartite entanglement constitutes one of the key resources in quantum information processing. We exploit correlation tensor norms to develop a framework for its experimental detection without the need for shared frames of reference. By bounding these norms for partially separable states and states of limited dimension we achieve an extensive characterization of entanglement in multipartite systems in an experimentally feasible way. Furthermore we show that both bi- and multipartite dimensionality of entanglement can be revealed by our methods.
\end{abstract}
\maketitle
\section{Introduction}
Multipartite entanglement appears to be the paradigmatic resource behind numerous quantum algorithms \cite{algo} and the very speed up of quantum computation itself \cite{vidal,jozsa}. While the exact role of this ubiquitous feature in quantum computing is still heavily debated \cite{vandenest,eisert}, much effort has been put into the characterization of entanglement in multipartite systems (see e.g. \cite{julio,gour,kraus,siewert}) and its experimental verification \cite{entdet}.\\
It is known that the complexity of faithfully telling whether a given quantum state is entangled is a hard problem, unlikely to yield an efficiently computable solution \cite{gurvits}. Furthermore the number of measurements required to fully characterize a given quantum state scales exponentially with the number of systems. Consequently most effort in the detection of multipartite entanglement has been directed towards entanglement witnesses or similar sufficient criteria for revealing entanglement \cite{entdet,siewert}. While being very efficient in terms of required measurements, these witnesses require a good agreement on a shared frame of reference to be successful.\\
In this letter we want to address these challenges by showing that local unitary invariant norms of correlation tensors can be used to reveal even the strongest form and the dimensionality of entanglement in multipartite systems. While these norms would require an exponential amount of measurements, they can be efficiently lower bounded just by performing a selected subset of measurements. This hybrid approach yields efficient entanglement witnesses that do not rely on shared reference frames. They rely only on correlation between observables in separated parties and are built in a sequential way, such that one can collect data until a sufficient number is reached and entanglement verified. Revealing entanglement from correlation tensor norms has previously been studied in the bipartite case \cite{julio1,julio2,badziag,lawson}, used for determining non-full separability in multipartite states \cite{john,hassan,julio3} and shown to be able to reveal multipartite entanglement in tri- and four- partite systems \cite{julio3}. In this manuscript we present the first successful detection of multipartite entanglement for any number of systems and the first detection of (multipartite) entanglement dimensionality using correlation tensor norms. { These results show that one can conclude genuine multipartite entanglement directly from local unitary invariant purity distributions of the multipartite states.}\\
To embark on that endeavor let us first define the relevant concepts. 

\section{DEFINITIONS AND NOTATION}

Pure states (i.e. projectors) are defined to be $k$-separable if they can be written as a $k$-fold tensor product $|\Phi\rangle\langle\Phi|_{k-sep}=|\phi_1\rangle\langle\phi_1|\otimes|\phi_2\rangle\langle\phi_2|\otimes\cdots\otimes|\phi_k\rangle\langle\phi_k|$. Quantum states that can be decomposed into $k$-separable projectors, i.e. $\rho_{k-sep}=\sum_ip_i|\Phi_i\rangle\langle\Phi_i|_{k-sep}$, are called $k$-separable themselves as they can be produced from sharing $k$-separable states and performing local operations aided by classical communication (LOCC). The strongest form of entanglement in multi partite systems is thus given by states which are not even $2$-separable and they are commonly referred to as genuinely multi partite entangled (GME).\\ 

\subsection{Correlation tensor and Bloch vector decomposition}

The Bloch vector decomposition offers a convenient way to write a density matrix in terms of expectation values and correlations of local observables. We can express a general $n$ partite qudit state as
\begin{align}
\rho=\frac{1}{d^n}\sum_{i_1,i_2,\cdots,i_{n}=0}^{d-1}T_{i_1,i_2,\cdots,i_{n}}\bigotimes_{k=1}^n\lambda_{i_k},
\end{align}
where we have used suitably normalized generators of the SU(d) that fulfill
\begin{align}
\text{Tr}[\lambda_{i}\lambda_{j}]=d\delta_{ij}\,,
\end{align}
and $\lambda_0=\mathbbm{1}_d$. It follows that the tensor elements are simply given by
\begin{align}
T_{i_1,i_2,\cdots,i_{n}}=\text{Tr}[\rho\bigotimes_{k=1}^n\lambda_{i_k}]\,.
\end{align}

\subsection{Lower order correlation tensors}

The state of a subsystem $\rho_\alpha=\text{Tr}_{\overline{\alpha}}[\rho]$ containing systems $\alpha\subseteq\{1,2,\cdots,n\}$ is completely determined by the tensor elements where all indices that are not part of $\alpha$ are set to zero. This yields a natural division of $T$ into subsystem correlation tensors, that encode correlations between all nontrivial observables within $\alpha$, i.e. $T=\sum_{\alpha\subseteq\{1,2,\cdots,n\}}\tau_\alpha$. We will focus our attention on the the full-body tensors $\tau_{1,2,\cdots,n}$, henceforth denoted simply as $\tau$. The regular $2$-norm of these tensors is given as
\begin{align}
||\tau(\rho)||:=\sqrt{\sum_{i_1,i_2,\cdots,i_{n}=1}^{d-1}T_{i_1,i_2,\cdots,i_{n}}^2}\,,
\end{align}
where the summation index now starts with $1$ to exclude all elements referring to identities in any number of subsystems.
Using our convention to label the full body correlation tensor norms of subsystems is thus straightforward: 
\begin{align}
||\tau_\alpha(\rho)||:=||\tau(\text{Tr}_{\overline{\alpha}}[\rho])||
\end{align}
We have chosen the $2$-norm of correlation tensors as it carries the advantage that it can be efficiently lower bounded by looking only at a subset of correlations. 

\section{Results}

Our first main result concerns the detection of non-$k$-separability from the $2$-norm of the full body correlations alone. 
\begin{theorem} For all $k$-separable states the $2$-norm of the full body correlation tensor is bounded from above by 
\begin{align}
||\tau(\rho_{k-sep})|| 
\leq \sqrt{(d^2-1)^{k}(d^{\lceil \frac{n}{k} \rceil-2})^{R}(d^{\lfloor \frac{n}{k} \rfloor-2})^{k-R}} \label{Equipartitionbound1}
\end{align}
with $R=n-k\lfloor \frac{n}{k} \rfloor$.
\end{theorem}
The proof is rather straightforward and can be summarized in a few words (the detailed derivation is presented in the supplemental material for the readers convenience): Since the $2$-norm of the correlation tensor is convex in the space of density matrices it is sufficient to bound the maximal norm for a $k$-separable pure state. For those the norms are multiplicative under tensor products, so one only needs to find the maximal $2$-norm of general states and then find the maximal product among all possible $k$-partitions of $n$.\\
These bounds enable reliable and robust detection of entanglement for general $n$ and even detection of GME for $n=3$. However most multipartite entangled states can not be revealed from full-body correlations alone and indeed this theorem fails at providing a detection criterion for verifying GHZ, Cluster or W-state type entanglement for $n>3$. Hence, when using the correlation tensor norm for the case $k=2$, i.e. the detection of GME, we need to include lower order correlation terms as well. For that purpose we adopt the short-hand notation $\mathcal{C}_x(\rho):=\sum_{m=x}^{n} \sum_{|\alpha| =m} ||\tau_{\alpha}||^2$ and find 
\begin{theorem}
For states separable for a given (bi-)partition $(k_1|k_2):k_1+k_2=n$, the sum of all squared $m$-body correlation tensor norms from $x$ to $n$ is bounded by
\begin{align}
\mathcal{C}_x(\rho_{2-sep})\leq (d^{k_{1}}-1)(d^{k_{2}}-1)+  \sum_{k_{j} \geq x}(d^{k_{j}}-1) \label{Cutoffthm}
\end{align}
\end{theorem} 
The basic proof behind this theorem makes use of the same concepts as before, however now deciding which of the bipartition bounds is the largest depends on the cutoff (or inclusion) parameter $x$ (More details can again be found in the supplemental material). In principle simple combinatorial considerations are sufficient to calculate the bounds for any $x$, so we now focus on two particularly useful cases.

\section{APPLICATION: ENTANGLEMENT DETECTION}

\subsection{The absolutely maximally entangled (AME) state}

\begin{cor}
For biseparable states (i.e. $k=2$), the sum of all squared $m$-body correlation tensor norms from $\frac{n}{2}+1$ to $n$ is bounded by
\begin{align}
\mathcal{C}_{\frac{n}{2}+1}(\rho_{2-sep}) \leq d^n-d \label{Cutoffthm2}
\end{align}
\end{cor} 
Using corollary 1 we can already state one of the main implications of the bounds: For every number of parties $n$ there exists a GME state that is detected by eq(\ref{Cutoffthm}).\\
As an exemplary case we can point to the maximally multipartite entangled state (MMES) \cite{MMES}, also known as absolutely maximally entangled state (AME) \cite{AME1,AME2,AME3}. These are states, whose marginals (i.e. reduced density matrices) are maximally mixed for all reductions smaller or equal to $\frac{n}{2}$. Using that $\text{Tr}(\rho^2)=\frac{1}{d^n}(\sum_{m=0}^{n} \sum_{|\alpha| =m} |\tau_{\alpha}|^2)=1$, this directly implies that
\begin{align}
\mathcal{C}_{\frac{n}{2}+1}(\rho_{AME})=\sum_{m=\frac{n}{2}+1}^{n} \sum_{|\alpha| =m} |\tau_{\alpha}|^2=d^n-1\,,
\end{align}
and thus the AME states violate the bi-separable bound for every $n$. Despite the inclusion of numerous correlation tensors, to violate the bound only $d^n-d+1$ correlations need to be ascertained, which is still less than the square root of a state tomography (which would require $d^{2n}-1$ correlation terms).\\

\subsection{Detecting non-biseparability}

Another case which turns out to be very useful is the case of including just the $n-1$-body correlations to the full body correlation tensor. Here we find
\begin{cor}
For bi-separable states the sum of all squared $m$-body correlation tensor norms from $n-1$ to $n$ is bounded by
\begin{align}
\mathcal{C}_{n-1}(\rho_{2-sep}) \leq \max[A,B] \label{Cutoffthm3}\,,
\end{align}
where $A=d^n-2d^{\frac{n}{2}}+1$ and $B=(d-1)(d^{n-1}-1)+d^{n-1}-1-\frac{n}{n-1}(d^{n-3}-1)$.
\end{cor} 

\subsection{Graph states}

Now that we have established bounds on correlations for different types of separable states we proceed by illustrating our method for exemplary cases and discuss how one would apply these criteria in a typical experimental scenario.\\
We want to start this discussion using a highly relevant state in measurement based quantum computation. For sake of simplicity let's commence with the first non-trivial case of four qubits and a qubit cluster state on a square lattice. First we start by simply computing the norms of the correlation tensors one would expect to be present when trying to create this state in a laboratory. Indeed for all stabilizer states this proves to be very easy, as the only non-zero correlation tensor elements in $m$-body correlation tensors correspond to $m$-body stabilizers and each have a value of $1$. This facilitates the computation of all tensor norms and in our case yields the following: $||\tau_{1,2,3,4}||^2=5$, $||\tau_{1,2,3}||^2=2$, $||\tau_{1,2,4}||^2=2$, $||\tau_{1,3,4}||^2=2$ and $||\tau_{2,3,4}||^2=2$. A quick check with eq.(\ref{Equipartitionbound1}) reveals that even if one managed to perfectly engineer the state, the bi-separable bound for the full-body correlation tensor is $9$, such that it would not be detected. Using eq.(\ref{Cutoffthm3}) and including the $3$-body correlations we find that the bound now reaches $37/3$. So summing up the $3$-body correlation tensor norms and including the full body correlations we reach a value of $13$. This implies that we can detect this state to be multipartite entangled from local unitary invariants alone (albeit with a rather small noise robustness in this case). Instead of measuring the full three and four body correlations one can concentrate on exactly the elements which according to the initial calculation should be non-zero. As mentioned before any element ascertained yields a lower bound on the norm of the tensor (as the full norm is the sum of the squared elements). In case of an imperfect alignment of reference frames there is a chance that expected correlations are not found and one needs to continue measuring element by element until the lower bound exceeds the separable bound and thus proves entanglement.\\

\subsection{Noise resistance}

\begin{figure}[tp!]
\begin{center}
  \includegraphics[width=0.5\textwidth]{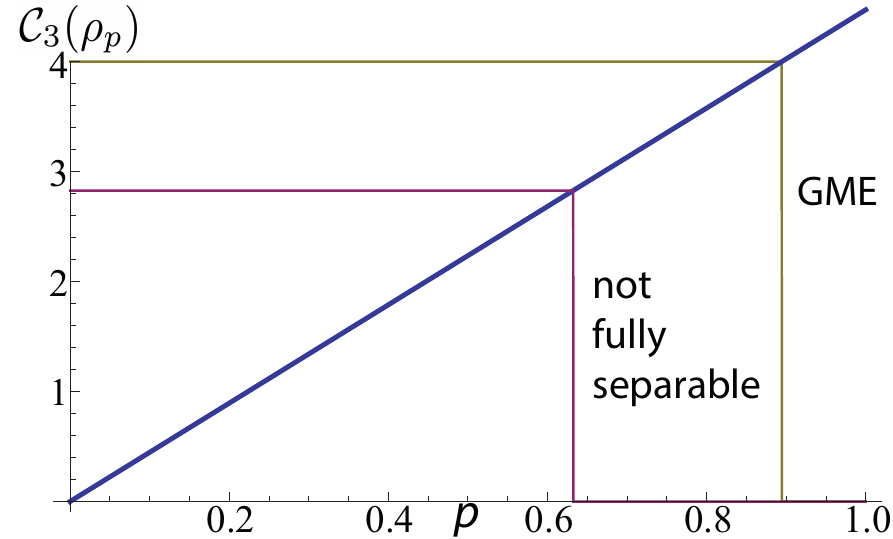}
  \caption{\label{Fig1} Here we illustrate the noise resistance of the entanglement detection criterion derived from eq.\ref{Cutoffthm3}. For a value of $p>\sqrt{\frac{2}{5}}$ the three dimensional generalized GHZ state mixed with white noise is detected to be entangled and for $p>\sqrt{\frac{4}{5}}$ it is detected to be genuinely multipartite entangled. { Compared to the noise resistance of witnesses directly tailored towards detecting exactly the this state, the exhibited noise resistance is of course rather weak (e.g. compared to $p>\frac{1}{4}$ from Ref.~\cite{julio}), which is the price to be paid for local unitary invariance. On the other hand the witnesses designed to detect exactly this state would inevitably fail if the state undergoes a local unitary rotation in just one of the subsystems.}}
\end{center}
\end{figure}

A typical benchmark in testing entanglement criteria is illustrating the resistance to the worst kind of possible noise. Since the criterion relies only on correlations it is obvious that the noise that is most detrimental to tensor norm type criteria is clearly a state without any correlation whatsoever, i.e. white noise. In Fig.\ref{Fig1} we illustrate the criterion for $x=2$ and $\rho_p=p*|GHZ_3\rangle\langle GHZ_3|+\frac{1-p}{27}\mathbbm{1}_{27}$, where $|GHZ_3\rangle=\frac{1}{\sqrt{3}}(|000\rangle+|111\rangle+|222\rangle)$ is the three dimensional generalization of the Greenberger-Horne-Zeilinger state.\\

\section{APPLICATION: DIMENSIONALITY WITNESS}

Beyond merely detecting whether a given quantum state is entangled, there has been some recent interest in describing features of entangled states relating to the necessary dimension needed for producing the correlations of a given state. For bipartite systems this entanglement dimensionality is known as Schmidt number \cite{sn}, is still subject to research \cite{chrisjens} and recently many experiments prove the capability of producing high dimensionally entangled state in the lab \cite{dimexp,dimexp2}.\\
In multipartite systems one can collect all necessary local ranks in a rank vector \cite{CHLW} and classify multipartite entanglement according to the required rank vectors in the decomposition \cite{julio}. A special case of this is given by the dimensionality of multipartite entanglement \cite{multidim}, defined as the minimum Schmidt rank across every bipartition, which can also be extended via convex roof constructions.\\
We continue by proving also a relation between correlation tensor norms and the dimensionality of entanglement, both in the bipartite and multipartite case. Here the local unitary invariance of the tensor norms facilitates the construction of entanglement dimensionality witnesses. For multipartite systems we present the following theorem
\begin{theorem}
For multipartite entangled states that can be decomposed into pure states with local ranks of $(k_1,k_2,\cdots,k_n)$, the correlation tensor can be bounded by
\begin{align}
\mathcal{C}_2(\rho_{(k_1,k_2,\cdots,k_n)}) \leq d^n+n-1-\sum_i\frac{d}{k_i} \label{Dimthm}
\end{align}
\end{theorem} 
This theorem is a consequence of the fact that any pure state with a bounded local Schmidt rank has nonzero local tensor norms $||\tau_i||^2\geq \frac{d}{k_i}-1$. This theorem can be readily generalized to also include marginals of more than one system (required to ascertain genuine multipartite entanglement dimensionality of systems beyond $n=3$) and enables the detection of the dimensionality structure of multipartite entanglement. As an application let us consider the question whether a given tripartite state $|\psi\rangle$ can be decomposed into multipartite entangled states with a Schmidt (or dimensionality) vector of $(2,2,3)$. The canonical overlap witness construction from Ref.~\cite{entdet} would yield a witness for this set given by
\begin{align}
W_{\overline{(2,2,3)}}:=\alpha\mathbbm{1}-|\psi\rangle\langle\psi|\,,
\end{align}
with $\alpha:=\max_{|\phi\rangle\in(2,2,3)}|\langle\phi|\psi\rangle|^2$. There is no known method how to calculate such $\alpha$, as the problem is a generalization of calculating a variant of the geometric measure of entanglement \cite{GM}. One way to detect such states through a different approach was presented in Refs.\cite{julio,MAMAJU}, requiring exact knowledge of the local observables. With our criterion we can immediately detect that the state $|\psi_1\rangle:=\frac{1}{\sqrt{6}}(|000\rangle+|111\rangle+|012\rangle+|102\rangle+|120\rangle+|021\rangle)$, which belongs to $(2,3,3)$, is not in $(2,2,3)$, as $\mathcal{C}_2[|\psi_1\rangle\langle\psi_1|]=25.5$ and from eq.(\ref{Dimthm}) we know that the bound for $(2,2,3)$ states is $25$.\\

\section{EXPERIMENTAL REALIZATION}

{Before concluding let us briefly discuss how an experimental estimation of the correlation tensor norms would look in practice. First one would measure in one basis with $d$ outcomes locally. These measurement results can be used to define the diagonal part of the generators of the $SU(d)$ and record all corresponding correlation elements with just one local setting. This can potentially already reveal some information about the achieved alignment, but instead of continuing to align one would proceed to measure dichotomic observables that are made up from superpositions of two eigenstates of the first local basis. With each measured correlation the norm can be updated (as the sum over all squared elements can easily be lower bounded by a partial sum) until the corresponding separability bound is violated and entanglement is certified. A natural candidate for such a procedure are e.g. photonic orbital angular momentum entangled states (as e.g. in \cite{dimexp2}). Here one could first locally measure Laguerre-Gauss (LG) \cite{LG} modes with a mode sorter \cite{MS} and then proceed to use spatial light modulators (SLM) for the two-dimensional subspace measurements that provide all necessary values for the symmetric and anti-symmetric $\lambda_i$.}\\

\section{CONCLUSION}

The analysis of our bounds on correlation tensor norms for different sets of states, has shown that these experimentally readily accessible norms can be used to characterize many important features of multipartite quantum states. They can reveal whether a given state in an experimental setup can be produced via $k$-separable states from LOCC, reveal genuine multipartite entanglement and even completely characterize entanglement dimensionality in multipartite systems, where the canonical approaches fail. Because some of the bounds obtained are rather crude and surely leave room for improvement we hope that this letter sparks renewed efforts into characterizing correlation tensors under different separability/dimension constraints.\\

{ The experimental advantages of such approaches manifest when a perfect alignment of reference frames is not possible, which especially for many parties at distant locations can provide a substantial advantage. To be experimentally feasible, however, the correlations should still be concentrated in as few elements as possible. If the correlations were spread thinly such that one would require almost a full state tomography one could just use different criteria that make use of the full density matrix, without sacrificing noise resistance for local unitary invariance.} Here one could study the performance of randomly chosen measurements and compare the results to the probabilistically reference frame alignment free approaches to non-locality from Refs.~\cite{rnl1,rnl2,rnl3}. To perform even better one should exploit that there are simple rules one could follow in order to exclude further measurements once significant correlation values are found. This sequential measurement scheme allows for process optimization. For example, once it is ascertained that the correlation between $\langle\lambda_i\otimes\lambda_j\rangle\approx d-1$, one can already exclude all measurements of $\langle\lambda_i\otimes\lambda_j'\rangle$ and $\langle\lambda_i'\otimes\lambda_j\rangle$. An open challenge is the development of more sophisticated techniques, that improve the efficiency of sequential correlation tensor acquisition.\\
\textit{Acknowledgements} We thank Steve Brierly, Cecilia Lancien and Andreas Winter for helpful discussions. MH acknowledges funding from the Juan de la Cierva fellowship (JCI 2012-14155), the European Commission (STREP "RAQUEL") and the Spanish MINECO Project No. FIS2013-40627-P, the Generalitat de Catalunya CIRIT Project No. 2014 SGR 966, CK acknowledges funding from the ERC Advanced Grant "IRQUAT".

\begin{widetext}
\section{Appendix}\label{Appendix}
\subsection{Bounding the Norm  of a Single Fullbody Correlation Tensor: Derivation}

Our first approach is focusing on the full body correlation tensor $\tau$. We aim to find an upper bound  of this tensor the 2-norm and for partially separable states. In particular we want to solve
\begin{align}
\max_{\sigma\in k-SEP}||\tau(\sigma)||\,,
\end{align}
where $k-SEP$ denotes the set of $k$-separable states. Due to the convexity of the norm it is sufficient to look at the extremal elements of the set, i.e. pure states. Furthermore, due to the multiplicativity of the norms of product states, the above optimization boils down to
\begin{align}
\max_{\sigma\in k-SEP}||\tau(\sigma)||=\max_{\{\tau_{\alpha_i}\}|(\dot{\bigcup}_i\alpha_i)=\{1,2,\cdots,n\}}||\tau_{\alpha_1}||\times||\tau_{\alpha_2}||\times\cdots\times||\tau_{\alpha_k}||\,.\label{bound}
\end{align}
We are interested in the achievable unconstrained maximum of the full body correlation tensors for any number of particles. 
Before deriving any bounds we want to make a simple yet important observation. 

\begin{obs2}
We express $\rho$ in the basis of generalized Pauli operators obtaining
\begin{align}
\Tr(\rho^{2})=\frac{1}{d^n}(1+||T||^2)=\frac{1}{d^n}(1+\sum_{\alpha}||\tau_{\alpha}||^2), \label{quadratictracerelation}
\end{align}  
\end{obs2}
This observation yields a direct connection between the trace of the squared density matrix and the correlation tensor norm, that will prove to be very useful. A simple upper bound on the 2-norm can be found via our observation and the fact that the tensor norms are again maximized by pure states, implying $Tr[\rho^2]=1$ and in turn
\begin{align}
\frac{1}{d^n}(\sum_\alpha||\tau_\alpha||^2+1)=1\label{Puritybound}
\end{align}
and thus $||\tau||\leq \sqrt{(d^{n}-1)}$. That this is already useful can be seen e.g. for three qubits, where the bound for biseparable states is $\sqrt{3}$ and the GHZ state has a norm of $2$ and is thus detected to be genuinely multi partite entangled.\\

\subsubsection{Bound for Equipartitions}

We establish a lemma, that will be used constantly through the text. In essence it describes for any $k$-partition of a set, which partition maximizes the norm, depending only on a few basic properties of the upper bounds on the individual norms. Generically such multiplicative bounds favor equipartitions:

\begin{lemma2}

If we have an upper bound $f(|\beta|)$ on $||\tau_{\beta}||$ depending on the cardinality $|\beta|$ of the multi index $\beta$, $y \in \mathds{N}$ that satisfies $f(|\beta|)^{2}\geq f(|\beta|+y)f(|\beta|-y)$
and a $k$ separable state, we can bound norms of the $k$ partitions correlation tensor $\beta_{i}$ by  

\begin{align}
||\tau_{\beta_1}||\times||\tau_{\beta_2}||\times\cdots\times||\tau_{\beta_k}||\, 
\leq \sqrt{(f(\lceil \frac{n}{k} \rceil))^{R}(f(\lfloor \frac{n}{k} \rfloor}))^{k-R}, \label{GeneralEquipartitionbound}
\end{align}
with $R=n-k\lfloor \frac{n}{k} \rfloor$.
\end{lemma2}

\begin{proof}
Let $|\beta|=x$,$|\beta'|=x+y$,$|\beta''|=x-y$ and  $\gamma$ arbitrary large for $x,y \in \mathds{N}$.

Then for any function of a tensor norm with $f(x)^{2}\geq f(x+y)f(x-y)$ we trivially get
\[
\frac{||\tau_{\beta'}|| \times ||\tau_{\beta''}|| \times ||\tau_{\gamma}||}{||\tau_{\beta}|| \times ||\tau_{\beta}|| \times ||\tau_{\gamma}||} 
\leq \frac{f(x+y)f(x-y)||\tau_{\gamma}||}{f(x)^{2}||\tau_{\gamma}||}\leq 1.
\]
Therefore $k$ partitions of equal size yield the maximal value.

However if $n/k$ is not a natural number we may not be able to choose all partitions to be equal. 
Note that the above case includes all partitions where the difference in size between two partitions $||\beta'|-|\beta''||=2y$ is larger than one and we can apply the observation above. 

However we have yet to discuss the special case of the partition sizes differing by one.  All partitions of the largest partitioning, have to be either of size $\lfloor \frac{n}{k} \rfloor$ or $ \lceil \frac{n}{k} \rceil$, else we could apply the above argument. $R=n-k\lfloor\frac{n}{k}\rfloor$ is the remainder of the division of n by k. In order to make sure all partitions sum up to $n$ exactly $R$ can be larger by one element than the remaining ones, proving our claim.
\end{proof}
The purity bound is by no means the only possible function that we can insert above for $f$, we will provide several different sensible choices later in the text, still we start by using (\ref{Puritybound}) as the simplest choice obtaining our first corollary.

\begin{cor2}
If we have a $k$ separable state, we can bound norms of the $k$ partitions correlation tensor $\beta_{i}$ by (\ref{Puritybound}) obtaining
\begin{align}
||\tau_{\beta_1}||\times||\tau_{\beta_2}||\times\cdots\times||\tau_{\beta_k}||\, 
\leq \sqrt{(d^{\lceil \frac{n}{k} \rceil}-1)^{R}(d^{\lfloor \frac{n}{k} \rfloor} - 1)^{k-R}} \label{Equipartitionbound1}
\end{align}
with $R=n-k\lfloor \frac{n}{k} \rfloor$.
\end{cor2}
\begin{proof}
The corollary follows from (\ref{Equipartitionbound1}) by using the purity bound (\ref{Puritybound}) $f(x)=d^{x}-1$. We calculate $f(x)^{2}=d^{2x}-d^{x}+1\geq d^{2x}-d^{x+y}-d^{x-y}+1=f(x+y)f(x-y)$ thus checking that our condition on $f$ is fulfilled.
\end{proof}

The upper bound above is derived only using the assumption of purity and $k$-separability. 
We emphasize that the norms above are convex and the maximum is therefore taken at the boundary of the set of states, the pure states. Any mixed state lies in the interior of the set of states and could only achieve smaller values. This means that if a states full body correlation tensor violates the above inequality, this state can not be $k$-separable.

\subsubsection{Improving the Purity-bound}

The bound given is not tight, as this would require pure states with all correlations concentrated in the full-body correlation tensor. That this is not possible is a simple consequence of the fact that the entropy of different partitions of pure states need to be equal and thus the higher order correlations below $n$, can generically not all be zero. Building upon this simple argument we try to improve our bound in this section by modifying (\ref{Puritybound}),
\begin{align}
||\tau||^2=d^{n}-1-\sum_{|\beta|<|\alpha|}||\tau_{\beta}||^{2}. \label{Puritybound2}
\end{align}
Any lower order correlation tensor element that has non zero norm will allow us to subtract something from $d^{n}-1$. 
At least for the $n-1$ body correlation tensor we can say something non trivial by using the Schmidt decomposition.

We establish a first improved bound out of the purity condition.
\begin{theorem2}
For the fullbody correlation tensor $\tau$ of a state $\rho$, 
\begin{align}
||\tau||^2\leq d^{n-2}(d^2-1).\label{Improvedbound}
\end{align}
\end{theorem2}

\begin{proof}
As a consequence of the Schmidt decomposition, any two bipartitions of a pure state have the same entropy, regardless of the entropy we use.
In our case it proves to be helpful to consider the special case of linear entropy yielding
\begin{align}
2(1+\Tr(\rho^{2}_{\alpha / j}))=S_{L}(\rho_{\alpha / j})=S_{L}(\rho_{j})=2(1+\Tr(\rho^{2}_{j})).
\end{align}
The choice of the linear entropy is motivated by the fact that a correlation tensor can be expressed in terms of $\Tr(\rho^{2})=\frac{1}{d}(1+||T||^{2})$ due to 
(\ref{quadratictracerelation}), allowing us to write out the condition that $\rho$ is a pure state
\begin{align}
\frac{1}{d}\leq\Tr(\rho_{j}^{2})=\Tr(\rho_{\alpha / j}^{2})=\frac{1}{d^{n-1}}(\sum_{\beta\subseteq\{1,\cdots,n\}/j}||\tau_{\beta}||^2+1),
\end{align}
or rewriting the above to bound the lower order correlation tensor elements
\begin{align} 
\sum_{\beta\subseteq\{1,\cdots,n\}/j}||\tau_{\beta}||^2 \geq d^{n-2} -1. \label{Lowertermbound}
\end{align}
We insert in (\ref{Puritybound2}) obtaining $||\tau||^2\leq d^{n-2}(d^2-1)$.
\end{proof}
This stronger bound can be inserted in the proof of the equipartition bound (\ref{GeneralEquipartitionbound}) to obtain a stronger result.
\begin{cor2}
If we have a $k$ separable state, we can bound norms of the $k$ partitions correlation tensor $\beta_{i}$ by (\ref{Puritybound}) obtaining
\begin{align}
||\tau_{\beta_1}||\times||\tau_{\beta_2}||\times\cdots\times||\tau_{\beta_k}||\, 
\leq \sqrt{(d^2-1)^{k}(d^{\lceil \frac{n}{k} \rceil-2}-1)^{R}(d^{\lfloor \frac{n}{k} \rfloor-2}-1)^{k-R}} \label{Equipartitionbound2}
\end{align}
with $R=n-k\lfloor \frac{n}{k} \rfloor$.
\end{cor2}
\begin{proof}
The corollary follows from (\ref{Equipartitionbound1}) by using (\ref{Improvedbound}). We calculate $f(n)^{2}=d^{2n-4}(d^{2}+1)^{2} = d^{2n+y-y-4}(d^{2}+1)^{2}=f(n+y)f(n-y)$ thus checking that our condition on $f$ is fulfilled.

\end{proof}

Using some basic combinatorial argument it is possible to further improve the bound:
\begin{theorem2}
For all non fullbody correlation tensors $\tau$ of states $\rho$,
\begin{align}
||\tau||^2 \leq d^n-\frac{n(d^{n-2}-1)}{(n-1)} 
\end{align}
\end{theorem2}

\begin{proof}
Applying (\ref{Lowertermbound}) we get 
\begin{align}
\sum_{j=1}^{n} \sum_{\beta\subseteq\{1,\cdots,n\}/j} ||\tau_{\beta}||^2 \geq n(d^{n-2}-1), \label{Improvedboundntimes}
\end{align}
we turn the above to into an improvement by observing
\begin{align}
\sum_{j} \sum_{\beta\subseteq\{1,2,\cdots,n\}/j}||\tau_{\beta}||^2=\sum_{|\beta|=n-1}||\tau_{\beta}||^2+\sum_{|\beta|=n-2}2||\tau_{\beta}||^2+(\cdots)+\sum_{|\beta|=1}(n-1)||\tau_{\beta}||^2. \label{Improvedboundlemma}
\end{align}
That the equality above holds can be seen by a short combinatorial argument. Some of the terms in $\sum_{\beta\subseteq\{1,2,\cdots,n\}/j}||\tau_{\beta}||^2$ will occur multiple times while summing over $j$. Let us consider the $i$-th coefficient $\tau_{\beta_{i}}$ with $|\beta_{i}|=k$, this index will occur once in every of the inner sums $\sum_{\beta\subseteq\{1,2,\cdots,n\}/j}||\tau_{\beta}||^2$ with $\beta_{i}\subseteq\beta$. To count the $\beta$ containing $\beta_{i}$ we fix the $k$ components equal to $\beta_{i}$ out of $|\{1,2,\cdots,n\}/j|=n-1$, leaving us to to pick $n-1-k$ elements. The components of $\beta$ can take $n$ different values, but they may not repeat themselves, by fixing $k$ of them we are left to choose from $n-k$ elements. Clearly this means that the an index of cardinality $k$ will occur $\binom{n-k}{n-k-1}=n-k$ times proving our observation. 

Combining (\ref{Improvedboundntimes}) and (\ref{Improvedboundlemma}) shows
\begin{align}
\frac{n(d^{n-2}-1)}{(n-1)} 
\leq \sum_{|\beta|=n-1} \frac{1}{(n-1)}||\tau_{\beta}||^2+\sum_{|\beta|=n-2}\frac{2}{(n-1)}||\tau_{\beta}||^2+(\cdots)+\sum_{|\beta|=1}||\tau_{\beta}||^2 
\leq \sum_{|\beta|<|\alpha|}||\tau_{\beta}||^2,
\end{align}
which by inserting in (\ref{Puritybound2}) proves our theorem.
\end{proof}
This is of course only an improvement for small $n$. Asymptotically it scales equal to the above bound in eq.(\ref{Improvedbound}).

In any case we have found two stronger bounds (\ref{Improvedbound}),(\ref{Improvedboundntimes}) than our initial purity bound (\ref{Puritybound}).
We can apply these instead of the weaker bound in (\ref{Equipartitionbound1}).

\subsection{Bounding the Norm  of Sums of Correlation Tensors: Derivation}

In the first part of the appendix, we have derived bounds for the full-body correlation tensor.
This serves for developing witnesses of states that have concentrated full-body correlations.
However concentrating all correlations in only the fullbody tensor, is a very extreme case of concentrated correlations and not achievable beyond $n=3$.
This motivates our second approach, where we derive bounds for states exhibiting most correlations in the "higher" tensors, but not only in a single one (For an example of such a state refer to the case of the AME state, which we discuss at the very end).
\begin{theorem2}For $0 \leq x < n$,the cutoff $ x \in \mathds{N}$,$\tau_{\alpha}$ the correlation tensor of $\rho$ , $\alpha$ an multi index denoting a $n$ partite biseparable (under the partition $\{\beta,\overline{\beta}\}$) qudit system, $|\beta|=k_{1}$ and $|\overline{\beta}|=k_{2}$ with $k_{1}+k_{2}=n$, the inequality
\begin{align}
\mathcal{C}_{x}(\rho):=\sum_{m=x}^{n} \sum_{|\alpha| =m} |\tau_{\alpha}|^2 = (d^{k_{1}}-1)(d^{k_{2}}-1) + \sum_{k_{j} \geq x}(d^{k_{j}}-1) \label{Cutoffthm}
\end{align}
holds.
\end{theorem2}

\begin{proof}
We split the sum of all correlation tensors into those containing at least an element of $\beta$ and $\overline{\beta}$ and those only containing elements from their respective partition. 
\begin{align}
\sum_{m=x}^{n} \sum_{|\alpha| =m} |\tau_{\alpha}|^2= \sum_{m=x}^{n}(\sum_{|\alpha| =m\wedge\alpha\subseteq\beta} |\tau_{\alpha}|^2 +\sum_{|\alpha| =m\wedge\alpha\subseteq\overline{\beta}} |\tau_{\alpha}|^2+\sum_{|\alpha| =m\wedge\alpha\nsubseteq\beta\wedge\alpha\nsubseteq\overline{\beta}} |\tau_{\alpha}|^2 )
\end{align}
We use the multiplicativity of the tensor norm, the purity (\ref{Puritybound}) and the biseparability bounding
\begin{align}
\sum_{m=x}^{n}\sum_{|\alpha| =m\wedge\alpha\nsubseteq\beta\wedge\alpha\nsubseteq\overline{\beta}} |\tau_{\alpha}|^2
\leq (d^{k_{1}}-1)(d^{k_{2}}-1).
\end{align}
However for the terms not containing an index of $\beta$ and $\overline{\beta}$, we can not make use of the separability.
We can still apply the purity bound. Take note that (\ref{Puritybound}) can be used for sums of tensors, therefore bounding the sum of a multi index and all other terms contained in the multi index by a single application of the purity bound. In each bipartition there is one maximal remainder term of cardinality $n-k_{i}$ containing all other terms. Therefore for each bipartition we get a bound
\begin{align}
\sum_{m=x}^{n}\sum_{|\alpha| =m\wedge\alpha\subseteq\beta} |\tau_{\alpha}|^2 \leq \begin{cases}(d^{k_{1}}-1)\,\text{for}\,k_1\geq x\\ 0\,\text{for}\,k_1< x\end{cases}\\
\sum_{m=x}^{n}\sum_{|\alpha| =m\wedge\alpha\subseteq\overline{\beta}} |\tau_{\alpha}|^2 \leq \begin{cases}(d^{k_{2}}-1)\,\text{for}\,k_2\geq x\\ 0\,\text{for}\,k_2< x\end{cases}
\end{align}
\end{proof}
If we use the trivial cut-off $x=0$ this results in
\[
\sum_{m=0}^{n}\sum_{|\alpha|} |\tau_{\alpha}|^2 \leq (d^{k_{1}}-1)(d^{k_{2}}-1) + (d^{k_{1}}-1) + (d^{k_{2}}-1) = d^{n} - 1
\]
that is unfortunately a trivial result for all pure states.
To avoid the trivial bound above we have considered the possibility of disregarding some of the lower order correlation tensors.
First we note that by (\ref{Puritybound}) we get for example $||\tau_{AB}||^{2}\leq (d^{2}-1)$ as well as for $||\tau_{AB}||^{2}+||\tau_{A}||^{2}+||\tau_{B}||^{2} \leq (d^{2}-1)$! The purity condition bounds a complete sum of correlation tensors, including all multi indices up to a chosen order, therefore the erasure of lower order correlation terms does not affect the purity bound. It is important to omit all terms containing a given index of the remainder term in order to achieve a reduction after application of the purity bound! For $k_{i}$ smaller than $x_{i}$ we simply omit the corresponding positive term $(d^{k_{i}}-1)$ obtaining our statement. 

We proceed with a quick discussion on the role of the cutoff parameter $x$ and it's relation to the size of the two partitions $k_{1}$,$k_{2}$.
We assume a bipartite state, therefore we only have to consider the three cases
\begin{itemize}
	\item $x<k_{1},k_{2}$: Since $x$ is smaller than any partition size and the the purity bound is invariant to the omission of lower order correlation tensor elements, we only arrive at the trivial purity bound $d^{n}-1$.
	\item $k_{1} \leq x<k_{2}$: 
	In this case (\ref{Cutoffthm}) reads $d^{n}-d^{k_{1}}$. The maximal value of this bound is achieved in the case of $k_{1}=1$ and $k_{2}=n-1$ yielding $d^{n}-d$.
	\item $k_{1},k_{2} \leq x$: (\ref{Cutoffthm}) reads $d^{n}-d^{k_{1}}-d^{k_{2}}+1$, this value is maximized whenever $|k_{1}-k_{2}|$ is minimal $d^{n}-d^{\left\lfloor \frac{n}{2} \right\rfloor}-d^{\left\lceil \frac{n}{2}\right\rceil}+1$.
	
\end{itemize}

\subsection{Bounds for Entanglement Dimensionality Detection}

In this paragraph we take a different viewpoint on (\ref{Cutoffthm}) and in order to use it as an entanglement dimensionality witness.

\begin{theorem2} 
For $n$-partite multipartite entangled states that can be decomposed into pure states with local $1$-party ranks \cite{CHLW} of $(k_1,k_2,\cdots,k_n)$
\begin{align}
\mathcal{C}_2(\rho_{(k_1,k_2,\cdots,k_n)}) \leq d^n+n-1-\sum_i(\frac{d}{k_i})
\end{align}
\end{theorem2} 

\begin{proof}

We write out $\mathcal{C}$
\begin{align}
\mathcal{C}_2(\rho_{(k_1,k_2,\cdots,k_n)})= \underbrace{ \sum_{m=2}^{n} \sum_{|\alpha| =m} |\tau_{\alpha}|^2 + \sum_{i} |\tau_{i}|^2}_{d^{n}-1}-
\sum_{i} \underbrace{|\tau_{i}|^2}_{\geq \frac{d}{k_i}-1}=d^n+n-1-\sum_i(\frac{d}{k_i}),
\end{align}
where we made use of the purity and the fact that any operator of dimension $k$ with positive eigenvalues and trace equal to one, can only have a squared trace as low as $\frac{1}{k}$.
\end{proof}

\subsection{The AME State}
	\subsubsection{Definition}
	\begin{definition}
	An \textbf{absolutely maximally entangled (AME) state}, is defined as a pure $n$-party state, with reduced density matrices equal to the maximally mixed state for every partition that is smaller or equal than $\frac{n}{2}$.
	\end{definition}
	\subsubsection{Detectability of the AME due to our bound}

	We know that for $|\alpha| \leq \frac{n}{2}$ the correlation tensor elements fulfill $||\tau_{\alpha}||=||\Tr(\lambda_{1}\otimes(\cdots)\otimes \lambda_{n} \frac{1}{d}\mathds{1})||=0$ by definition of the AME. The AME is pure, thus

	\begin{align}
	1=\Tr(\rho^{2}_{AME})=\frac{1}{d^{n}}(1+\sum_{m>\frac{n}{2}}\sum_{|\alpha|=m}||\tau_{\alpha}^{AME}||^{2}+\sum_{m\leq\frac{n}{2}}\sum_{|\alpha|=m}||\tau_{\alpha}^{AME}||^{2})=\frac{1}{d^{n}}(1+\sum_{m>\frac{n}{2}}\sum_{|\alpha|=m}||\tau_{\alpha}^{AME}||^{2}+0) \label{AMEcorr}
	\end{align}
	
shows that all nonzero correlations are the correlation tensor elements of size bigger than $\frac{n}{2}$ and $\sum_{m>\frac{n}{2}}\sum_{|\alpha|=m}||\tau_{\alpha}^{AME}||^{2}=d^{n}-1$. This illustrates why our cutoff theorem (\ref{Cutoffthm}) is sensible, setting $x=0$ results as well in the trivial bound  $d^{n}-1$ in (\ref{Cutoffthm}), since this coincides with the above we can not detect entanglement without an appropriately chosen $x$! However if $x$ is large enough \ref{Cutoffthm} provides a smaller bound than (\ref{AMEcorr}), allowing us to detect the AME! $x$ is large enough, whenever it is chosen larger than the size of the smallest partition. By choosing $x=\frac{n}{2}$ we can ensure that there exists a partition smaller or equal than $x$. 

In the discussion following (\ref{Cutoffthm}) we state the possible values this bound can achieve, depending on the size of $x,k_{1},k_{2}$. For $x=\frac{n}{2}$ the first case is excluded and the maximal remaining value that can be achieved is $d^{n}-d$ in the case $k_{1}=1<x<k_{n-1}$.
Comparing the correlation tensor norm of the AME state $d^{n}-1$ to the maximal correlation tensor norm achievable by a bipartite state $d^{n}-d$ we get a total separation of $d-1$, allowing for detection of the AME state.

\subsubsection{Full body correlation tensor of an AME}
We conclude by a short remark on the full body correlation tensor of the AME, thereby motivating what led us to derive the above (\ref{Cutoffthm}). The first bounds (\ref{Improvedbound}),(\ref{Improvedboundntimes}) we derived are suited for detection of states, whose correlation are concentrated in the full body part. Contrary to our initial hope, the AME  is not in general such a state since by the Schmidt decomposition we have 
\begin{align}
\frac{1}{d} = \Tr( (\rho_{j}^{AME})^{2})=\Tr((\rho_{\alpha / j}^{AME})^{2})=\frac{1}{d^{n-1}}(1+\sum_{\alpha / j} ||\tau_{\alpha}^{2}||),
\end{align}
illustrating $\sum_{\alpha / j} ||\tau_{\alpha}^{2}||=d^{n-2}-1$, due to this we know that many of the lower order correlation elements are non zero.
This makes it plausible that AME states are not in general detected by our bounds containing just  the full body correlation tensor (\ref{Improvedbound}),(\ref{Improvedboundntimes}), but only by bounds on sums of correlation tensor elements.
\end{widetext}

\end{document}